\documentstyle[amssymb,preprint,aps]{revtex}

\input{tcilatex}

\begin{document}
\title{Charge ordering and magneto-polarons in ${\bf Na_{0.82}CoO_{2}}$}
\author{C. Bernhard, A.V. Boris, N.N. Kovaleva, G. Khaliullin, A. Pimenov,
L. Yu, D.P. Chen, C.T. Lin, and B. Keimer}
\address{ Max-Planck-Institute for Solid State Research, Heisenbergstrasse
1, D-70569 Stuttgart, Germany.}
\date{\today}
\maketitle

\begin{abstract}
Using spectral ellipsometry, we have measured the dielectric function of a ${\rm %
Na_{0.82(2)}CoO_{2}}$ crystal that exhibits bulk antiferromagnetism with $%
{\rm T_{N}=19.8}$ K. We identify two prominent transitions as a
function of temperature. The first one at 280 K involves marked
changes of the electronic and the lattice response that are
indicative of charge ordering in the CoO$_{2}$ layers. The second
transition coincides with T$_{N}$=19.8 K and reveals a sizeable
spin-charge coupling. The data are discussed in terms of charge
ordering and formation of magneto-polarons due to a charge-induced
spin-state transition of adjacent Co$^{3+}$ ions.
\end{abstract}


The recent discovery \cite{Takada03} of superconductivity at
T$_{c}\simeq $5 \ K in the hydrated cobaltate ${\rm
Na_{0.35}CoO_{2}\ast 1.3\,H_{2}O}$ has engendered many proposals
for unusual electronic correlations. A particularly interesting
perspective is a spin-triplet pairing state, which has been
proposed on the basis of model calculations
\cite{Tanaka03,Singh03} and NMR experiments
\cite{Waki03,Fujimoto03}. In this context, the unusual
electromagnetic properties of the host material Na$_{x}$CoO$_{2}$
have also obtained renewed attention. Its layered crystal
structure gives rise to strongly anisotropic electronic properties
\cite{Terasaki97,Singh00}. The essential structural element are
metallic CoO$_{2}$ layers which consist of rhombohedrally
distorted, edge sharing ${\rm CoO_{6}}$ octahedra \cite{Neutrons}.
The triangular coordination of Co favors geometrically frustrated
and thus unconventional magnetic and electronic ground states. The
evidence includes an anomalous T-dependence of resistivity and
Hall-effect \cite{Wang03} and an unusually large thermo-electric
power in excess of 100 $\mu $V/K \cite{Terasaki97} that is
strongly suppressed in a magnetic field \cite{Wang03}.
Furthermore, angle resolved photo-emission spectroscopy (ARPES)
reveals an extremely narrow ($\sim $70 meV) and strongly
T-dependent quasi-particle band \cite{ARPES} in contrast to the
calculated band-width of about 1 eV \cite{Singh00}.

In an attempt to obtain additional information on the interplay of charge
and spin degrees of freedom, we have performed broad-band (80-44000 cm$^{-1}$%
) ellipsometry measurements of the in-plane and out-of-plane dielectric
function of a Na$_{0.82(2)}$CoO$_{2}$ single crystal that exhibits bulk
antiferromagnetic (AF) order below T$_{N}$=19.8 K \cite{Bayrakci03}. Our
data complement recent reports for samples with lower Na content of x=0.58 %
\cite{Lupi03} and 0.7 \cite{WangIR03} where static magnetism is
absent. We observe two prominent transitions as a function of T
that were previously not identified with optical experiments. The
first transition occurs below 280 K and involves sizeable changes
of the electronic and the lattice response that are suggestive of
polaron formation and charge ordering within the CoO$_{2}$ layers.
The second transition evident in the electronic response coincides
with T$_{N}$=19.8 K and indicates a strong spin-charge coupling.

A Na$_{0.82(2)}$CoO$_{2}$ single crystal (3x4x0.3 mm$^{3})$ was grown in an
optical floating zone furnace and characterized by x-ray, transport,
magnetic susceptibility, specific heat, and muon-spin rotation ($\mu $SR) %
\cite{Bayrakci03}. Susceptibility and $\mu $SR experiments established that
it exhibits bulk AF order below T$_{N}$=19.8 K \cite{Bayrakci03}. The
ellipsometric measurements were performed for the range of 0.01-0.8 eV with
a home-built ellipsometer in combination with a fast-Fourier-interferometer
(FTIR) at the IR-beamline of the ANKA synchrotron at FZ Karlsruhe, Germany %
\cite{Bernhard04}. Another home-built ellipsometer \cite{Kircher92} was used
for the range of 0.5-5.6 eV.

Figure 1 displays the in-plane dielectric properties between 8 and 300 K in
terms of the real parts of the optical conductivity, $\sigma _{1ab}$=1/4$\pi
\cdot \nu \cdot \varepsilon _{2ab},$ and the dielectric function, $%
\varepsilon _{1ab}$. Figure 1a shows $\sigma _{1,ab}$ for the entire range
from 80 to 44.000 cm$^{-1}$ at 25 and 300 K. The solid arrows mark four
bands corresponding to interband transition at 35850 cm$^{-1}$, 23850 cm$%
^{-1}$, 12200 cm$^{-1}$ , 8800 cm$^{-1}$ (1eV $\triangleq $ 8066 cm$^{-1}$).
Based on the band structure calculations \cite{Singh00} we assign the two
lower bands to transitions between Co 3d derived bands and the stronger high
frequency ones to charge transfer transitions between O-2p and Co-3d bands.

Figures 1b to 1e detail the frequency range below these interband
transitions. First we concentrate on the spectra at T
\mbox{$>$}
T$_{N}$=19.8 K. At 300 K the spectral shape of $\sigma _{1ab}$ is
characteristic of an incoherent transport mechanism. Whereas it is
almost constant between 600 and 6000 cm$%
^{-1}$, it decreases below 600 cm$^{-1}$ and
extrapolates well towards the dc value (black solid circle) \cite{Bayrakci03}%
. Below 300 K the optical spectra undergo marked changes. With
decreasing T a partial gap, a so-called pseudogap (PG), develops
which gives rise to a progressive decrease of $\sigma _{1ab}$
between 300 and 4500 cm$^{-1}$. Its onset is marked by an
absorption band at higher energies, which most likely corresponds
to charge excitations across the PG rather than to a conventional
interband transition. At 300 K this band is still fairly weak and
broad, whereas at lower T it rapidly sharpens, gains additional
spectral weight (SW), and
exhibits a sizeable blue-shift from about 3500 cm$^{-1}$ at 300 K to 4500 cm$%
^{-1}$ at 25 K. We argue below that the PG and the absorption band originate
from a partially charge ordered state and the excitations thereof.

As detailed in Figs. 1c and 1d, the PG formation is also accompanied by the
growth of two distinct low-frequency electronic modes. A band near 150 cm$%
^{-1}$ develops below 300 K.\ Its SW increases at low T, and the
simultaneous appearance of anomalous phonon modes (as discussed
below) suggest that it involves small polarons \cite{Calvani01}.
In the presence of a charge ordered state (as suggested by the PG
formation) it may also correspond to a pinned collective phase
mode \cite{Gruener85}. In addition we identify a very narrow
Drude-peak at the origin which accounts for the metallic
T-dependence of the dc conductivity \cite{Bayrakci03}. While the
narrow Drude-peak is not captured by the $\sigma _{1,ab}$ \
spectra at $\omega $%
\mbox{$>$}%
80 cm$^{-1}$, its presence is suggested by the dc conductivity which
increases rapidly below 300 K to $\sigma _{dc}\approx $5000 $\Omega ^{-1}$cm$%
^{-1}$
\mbox{$>$}%
\mbox{$>$}
$\sigma (\omega =80cm^{-1})\approx 6$00 $\Omega ^{-1}$cm$^{-1}$at 25K.
Moreover, the inductive free carrier response is apparent in $\varepsilon
_{1ab}$ for $\omega >$ 80 cm$^{-1}$ where a decrease towards negative values
at the origin is superimposed on the response of the polaronic mode at 150 cm%
$^{-1}$. Using a Drude-Lorentz-model we obtain at 25 K (dashed black line in
Fig. 1d) $\omega _{pl}=$1300(1) cm$^{-1}$ and $\gamma ^{D}=30(10)$ cm$^{-1}$
for the plasma frequency and the scattering rate of the free carriers,
respectively. The small plasma frequency corresponding to $\omega _{pl}^{2}=%
\frac{2e^{2}}{\pi }\frac{n}{m^{\ast }},$ with $n$ the carrier
density and $m^{\ast}$ their effective mass, is consistent with
the very narrow and weakly dispersive quasi-particle band as seen
in ARPES \cite{ARPES}. For the polaronic band we deduce an
oscillator strength S=450(50), a center
frequency $\omega $=150(1) cm$^{-1}$, and a full width at half maximum, $%
\gamma $=310(30) cm$^{-1}$.

Next we discuss the characteristic changes in the lattice response
which underline that the PG and the polaron band are signatures of
a charge ordering transition. Figures 1c, d and f highlight that
additional phonon modes suddenly appear below 300 K. Some of these
exhibit very asymmetric Fano line shapes that are indicative of a
strong interaction with the electronic background, in particular,
with the polaron band. At 300 K two phonon modes are resolved at
190 and 545 cm$^{-1}$ which are assigned to vibrations of Na
against O and Co against O, respectively \cite{Lemmens03}. In
addition there is a weaker mode at 515 cm$^{-1}$ which has not yet
been assigned. Figure 1f details the T-dependence around the
strongest mode near 550 cm$^{-1}$. Between 280 and 260 K this mode
suddenly splits into a broader mode around 545 cm$^{-1}$ and a
narrower one at 565 cm$^{-1}$. In addition the weaker mode around
515 cm$^{-1}$ seems to split into modes at 490 and 525 cm$^{-1}$.
Simultaneously, several narrow modes appear below 280 K at 585,
430, 390, 335, 300, 275, and 250 cm$^{-1}$. These sudden changes
in the lattice response highlight that a structural transition
takes place between 260 and 280 K. Interestingly, this transition
is not clearly resolved in the Raman-spectra nor in preliminary
x-ray diffraction measurements on small pieces from this crystal,
suggesting that the structural changes are fairly small. The
sizeable oscillator strength of the additional IR-active phonons
thus requires a charge density modulation which strongly couples
to the lattice distortions and thereby enhances their dipolar
moment. Overall, our optical data provide evidence that the charge
carriers within the CoO$_{2}$ layers exhibit a charge ordering
transition between 260 and 280 K. This conclusion is based on (i)
the large splitting of the CoO stretching mode and the
simultaneous formation of (ii) the PG, (iii) the polaron band, and
(iv) the narrow Drude-like response. A freezing transition of the
Na ions and subsequent small structural changes involving mainly
the Na layer may influence the charge ordering pattern, but cannot
account for the highly anomalous optical response. Notably, our
conclusion is supported by recent NMR experiments for x=0.75 where
a change in the NMR line shape and a sudden reduction in the
Co-1/T$_{1}$ relaxation rate slightly below 300 K were also
interpreted in terms of charge ordering in the CoO$_{2}$ planes \cite%
{Gavilano03}.

Next we discuss the second transition of the electronic response
which coincides with T$_{N}$=19.8 K. It is evident in Figs. 1c to
1d that the AF transition has a significant impact on the
electronic excitations including the PG and the polaron mode at
150 cm$^{-1}$. Apparently, it leads to a partial reversal of the
SW transfer that is associated with the charge ordering transition
below 280 K. A significant amount of SW is removed from the band
at 4500 cm$^{-1}$ and transferred to lower frequencies where it
partially fills in the PG. The polaron band at 150 cm$^{-1}$ also
exhibits a noticeable anomaly. A sizeable amount of SW is removed
from its center at 150 cm$^{-1}$ and redistributed towards higher
frequency. Concerning the Drude-peak we cannot draw a firm
conclusion. A small upturn of the dc resistivty below T$_{N}$
\cite{Bayrakci03} might indicate an anomalous SW loss of the
Drude-peak, but it may also be explained in terms of an enhanced
scattering from static magnetic defects. A reduction
of the free carrier response is not evident in the low frequency part of $%
\varepsilon _{1}$. This unresolved issue notwithstanding, our
optical data highlight a sizeable spin-charge coupling and suggest
the presence of magneto-polarons.

This conclusion is underlined by the c-axis optical response as
shown in Fig. 2 where the AF transition has an even stronger
impact on the polaronic band. At 300 K the electronic c-axis
response is rather weak and featureless. The spectra are dominated
by two IR-active phonon modes at 300 and 585 cm$^{-1}$ which are
assigned to out-of-plane vibrations of Na against O and Co against
O, respectively \cite{Lemmens03}. In analogy to the in-plane
response, the charge ordering transition below 280 K is
accompanied by the formation of a polaronic band. In contrast to
the in-plane response, however, there is no evidence for a PG
formation. To the contrary, the inset of Fig. 2b shows that the SW
nearly doubles between 300 and 25 K for the range below
3000 cm$^{-1}$. Accordingly, the electronic anisotropy between 80 and 4000 cm%
$^{-1}$ decreases from about 10 at 300\ K to less than three at 25 K. The
much larger anisotropy of $\sigma _{ab}^{dc}$/$\sigma _{c}^{dc}\sim $200 %
\cite{Terasaki97} is related to the absence of a Drude-response in
the c-axis component. Here the dominant feature is the polaronic
band centered around 400 cm$^{-1}$. Notably, it is subdivided by
several deep and narrow minima near 265, 335, 565 and 665
cm$^{-1}$ which can be accounted for in terms of extremely
asymmetric Fano-modes. These unusual anti-resonances may
correspond to defect modes or to local vibrational modes whose
response is out-of-phase with the one of the broad polaronic band.
The second prominent feature in the c-axis spectra is the strong
modification of the polaron band in the AF state. It becomes
strongly suppressed below 500 cm$^{-1}$ whereas its SW increases
at higher frequency up to 3000 cm$^{-1}$. This marked blue shift
of the polaronic band signals an increased self-trapping energy of
the magneto-polarons in the AF state where polaron hopping
requires spin-flip excitations.

In an attempt to interpret the unusual optical response of Na$_{0.82}$CoO$%
_{2}$ we note that strong polaronic features are well known in the
related compound La$_{1-y}$Sr$_{y}$CoO$_{3}$
\cite{Calvani01,Reik67}. Furthermore,
it has been suggested that a charge induced spin-state transition of the Co$%
^{3+}$ ions takes place in La$_{1-y}$Sr$_{y}$CoO$_{3}$ which gives rise to a
sizeable spin-charge coupling \cite{Yamaguchi96,Loshkareva01}. The
underlying idea is that a localized positive charge (corresponding to a Co$%
^{4+}$ ion) lowers the crystal field symmetry of adjacent
Co$^{3+}$ ions and thus significantly reduces the crystal field
splitting, $\Delta \left( e_{g}-t_{2g}\right) $, of the 3d levels
favoring an intermediate-spin (IS) state with S=1 over a low-spin
(LS) state with S=0. The coupled spin-charge object can be viewed
as a small magneto-polaron. The case of
dilute magneto-polarons applies to Na$_{x}$CoO$_{2}$ (La$_{1-y}$Sr$_{y}$CoO$%
_{3}$) for large x (small y). With decreasing Na-content (increasing y)
these magneto-polarons start to overlap and develop spatial correlations. In
the absence of additional disorder, a perfect triangular Wigner-lattice of Co%
$^{4+}$ ions could be realized at quarter filling, i.e. for x=0.75 \cite%
{Baskaran03}. A sketch is shown as an inset in Fig. 1e. In this state, the Co%
$^{3+}$ sites are arranged on a Kagom\'{e}-lattice and their IS state can be
stabilized due to the axial crystal field originating from the two
neighboring Co$^{4+}$ ions. Following these arguments, our present Na$%
_{0.82}$CoO$_{2}$ crystal should exhibit a charge ordered state
with a considerable number of defects. It is an interesting but
yet open question whether these defects tend to segregate and form
microscopic regions without charge order. Irrespective of the
details, this scenario accounts for the PG in terms of an
incomplete charge excitation gap. The large frequency scale of the
PG of about 4500 cm$^{-1}$ is explained by the gain in Hund
coupling, J$_{H}$, associated with the LS to IS\ transition. It
also provides a strong spin-charge coupling and therefore
naturally explains the large magneto-polaronic effects that occur
in the AF state. The IS state of Co$^{3+}$ is Jahn-Teller active,
which may explain the observed lattice anomalies. An interesting
question concerns the role of the charge disordered regions which
seem to give rise to the broad and incoherent electronic
background that dominates above the charge ordering transition,
i.e. above 280 K. At present we can not make an unambiguous
assignment of the narrow Drude-peak and the polaronic band at 150
cm$^{-1}$. They either arise from mobile or weakly localized
carriers within the charge disordered regions, or correspond to
unpinned or pinned collective phase modes within the charge
ordered regions. The latter possibility is favored based on recent
IR-reflectivity data on samples with lower Na content where this
kind of charge order should vanish. These data indicate indeed
that the PG and the low-frequency polaronic band are weakened at
x=0.7 \cite{WangIR03} and finally absent at x=0.58 \cite{Lupi03}.
FOr these Na concentrations, the Drude-peak is significantly
broader and less intense than the extremely narrow one in our
Na$_{0.82}$CoO$_{2}$ crystal. Finally, another interesting issue
concerns the role of the charge order induced S=1 spins of the
Co$^{3+}$ ions in the AF state. As they reside on a frustrated
Kagom\'{e}-lattice, they may well remain in a quantum liquid state
and not participate in the static magnetic order. Nevertheless,
they will mediate an in-plane exchange coupling (likely
ferromagnetic) between the S=1/2 moments of Co$^{4+}$ . The idea
that the charge ordered state is a prerequisite for static
magnetic order is indeed consistent with the finding that static
magnetism occurs only for x$\geqq $0.75 \cite{Sugiyama03}. The
expected spin and charge ordering scenario is also in good
agreement with the observation of three different
muon-spin-precession frequencies \cite{Bayrakci03}.

In summary, we have performed broad-band ellipsometric measurements of the
in-plane and c-axis dielectric function of a Na$_{0.82}$CoO$_{2}$ single
crystal that exhibits bulk AF order below T$_{N}$=19.8 K \cite{Bayrakci03}.
We identify two prominent transitions as a function of T. The first one
below 280 K involves changes of the electronic and the lattice response
which are suggestive of charge ordering in the CoO$_{2}$ layers. The most
prominent features are a pseudogap with an energy scale of about 4500 cm$%
^{-1}$ and a low frequency band which is assigned to small polarons. The
second transition coincides with T$_{N}$=19.8 K. It strongly affects the
polaronic band and the pseudogap and highlights a strong coupling between
electronic and magnetic degrees of freedom that is discussed in terms of
magneto-polarons, i.e. doped charges that stabilize an intermediate spin
state on neighbouring Co$^{3+}$ ions.

We gratefully acknowledge the support by Y.L. Mathis and B.
Gasharova at the ANKA synchrotron and valuable discussions with P.
Lemmens.

\begin{center}
\bigskip \newpage
\end{center}

Figure 1: \ The in-plane dielectric properties in terms of the
real parts of the optical conductivity, $\sigma _{1ab}$=1/4$\pi
\cdot \nu \cdot \varepsilon _{2ab},$ and the dielectric function,
$\varepsilon _{1ab}$. The spectra of $\sigma _{1ab}$ {\bf (a)} for
the entire spectral range at 300 and 25 K (arrows mark interband
transitions), {\bf (b)} below 6000 cm$^{-1}$ and {\bf (c)} below
650 cm$^{-1}$ (dc values from Ref. \cite{Bayrakci03} are shown by
solid circles, low-frequency extrapolations by dotted lines).
Corresponding spectra of $\varepsilon _{1ab}$ are shown in {\bf
(d)} and {\bf (e)}. The black dotted line in {\bf (d)} shows the
Drude-Lorentz fit at 25 K as outlined in the text. The inset of
{\bf (e)} gives a sketch of the suggested charge ordering pattern
at x=0.75. {\bf (f)} Splitting of the strongest phonon mode. The
dotted lines indicate the phonon splitting. Offsets have been
added to the spectra for clarity.

Figure 2: The c-axis optical conductivity, $\sigma _{1c}$, for polarization
perpendicular to the CoO$_{2}$ layers. {\bf (a)} Detailed view of the
low-frequency range below 650 cm$^{-1}$ and {\bf (b) }the full measured
range below 5000 cm$^{-1}$. The inset shows the integrated spectral weight
SW=$\int\limits_{80\ cm^{-1}}^{\omega }\sigma _{1c}\left( \omega ^{\prime
}\right) d\omega ^{\prime }$ at different T.

\end{document}